\documentclass[
 aps,prl,showpacs,superscriptaddress,numerical,amsmath,amssymb,floatfix,
reprint
]{revtex4-1}
\usepackage[dvipsnames]{xcolor}
\usepackage[
	pdffitwindow=true,
	colorlinks=true,
	frenchlinks=false,
        linkcolor=blue,
	anchorcolor=blue,
        citecolor=blue,
        filecolor=blue,
        urlcolor=blue,
        bookmarks=true,
        bookmarksopen=true,
	bookmarksnumbered=true,
        bookmarksopenlevel=1,
        plainpages=false,
	pdfpagelayout=TwoPageLeft,
        pdfpagelabels=true,
	breaklinks
]{hyperref}
\usepackage[per-mode=symbol,separate-uncertainty]{siunitx}
\usepackage{graphicx}
\usepackage{dcolumn}
\usepackage{color}
\usepackage{graphicx,wrapfig,lipsum}
\usepackage{bm}
\usepackage[english]{babel}
\usepackage{color}
\usepackage{ulem}
\usepackage{float}
\usepackage{chemformula}

\begin{document}
\title[]{Temperature-dependent spin-transport and current-induced torques in superconductor/ferromagnet heterostructures}

\author{M.~M\"uller}
\email[]{manuel.mueller@wmi.badw.de}
\affiliation{Walther-Mei{\ss}ner-Institut, Bayerische Akademie der Wissenschaften, 85748 Garching, Germany}
\affiliation{Physik-Department, Technische Universit\"{a}t M\"{u}nchen, 85748 Garching, Germany}

\author{L.~Liensberger}
\affiliation{Walther-Mei{\ss}ner-Institut, Bayerische Akademie der Wissenschaften, 85748 Garching, Germany}
\affiliation{Physik-Department, Technische Universit\"{a}t M\"{u}nchen, 85748 Garching, Germany}

\author{L.~Flacke}
\affiliation{Walther-Mei{\ss}ner-Institut, Bayerische Akademie der Wissenschaften, 85748 Garching, Germany}
\affiliation{Physik-Department, Technische Universit\"{a}t M\"{u}nchen, 85748 Garching, Germany}

\author{H.~Huebl}
\affiliation{Walther-Mei{\ss}ner-Institut, Bayerische Akademie der Wissenschaften, 85748 Garching, Germany}
\affiliation{Physik-Department, Technische Universit\"{a}t M\"{u}nchen, 85748 Garching, Germany}
\affiliation{Munich Center for Quantum Science and Technology (MCQST), Schellingstr. 4, 80799 M\"{u}nchen, Germany}
\author{A.~Kamra}
\affiliation{Center for Quantum Spintronics, Department of Physics, Norwegian University of Science and Technology, NO-7491 Trondheim, Norway}
\author{W.~Belzig}
\affiliation{Fachbereich Physik, Universit\"at Konstanz, 78457 Konstanz, Germany}
\author{R.~Gross}
\affiliation{Walther-Mei{\ss}ner-Institut, Bayerische Akademie der Wissenschaften, 85748 Garching, Germany}
\affiliation{Physik-Department, Technische Universit\"{a}t M\"{u}nchen, 85748 Garching, Germany}
\affiliation{Munich Center for Quantum Science and Technology (MCQST), Schellingstr. 4, 80799 M\"{u}nchen, Germany}
\author{M.~Weiler}
\affiliation{Walther-Mei{\ss}ner-Institut, Bayerische Akademie der Wissenschaften, 85748 Garching, Germany}
\affiliation{Physik-Department, Technische Universit\"{a}t M\"{u}nchen, 85748 Garching, Germany}
\author{M.~Althammer}
\email[]{matthias.althammer@wmi.badw.de}
\affiliation{Walther-Mei{\ss}ner-Institut, Bayerische Akademie der Wissenschaften, 85748 Garching, Germany}
\affiliation{Physik-Department, Technische Universit\"{a}t M\"{u}nchen, 85748 Garching, Germany}
\date{\today}
\begin{abstract}
We investigate the injection of quasiparticle spin currents into a superconductor via spin pumping from an adjacent FM layer.$\;$To this end, we use NbN/\ch{Ni80Fe20}(Py)-heterostructures with a Pt spin sink layer and excite ferromagnetic resonance in the Py-layer by placing the samples onto a coplanar waveguide (CPW). A phase sensitive detection of the microwave transmission signal is used to quantitatively extract the inductive coupling strength between sample and CPW, interpreted in terms of inverse current-induced torques, in our heterostructures as a function of temperature. Below the superconducting transition temperature $T_{\mathrm{c}}$, we observe a suppression of the damping-like torque generated in the Pt layer by the inverse spin Hall effect (iSHE), which can be understood by the changes in spin current transport in the superconducting NbN-layer. Moreover, below $T_{\mathrm{c}}$ we find a large field-like current-induced torque.
\end{abstract}
\maketitle
\begin{figure}
	\centerline{\includegraphics[scale=0.265625]{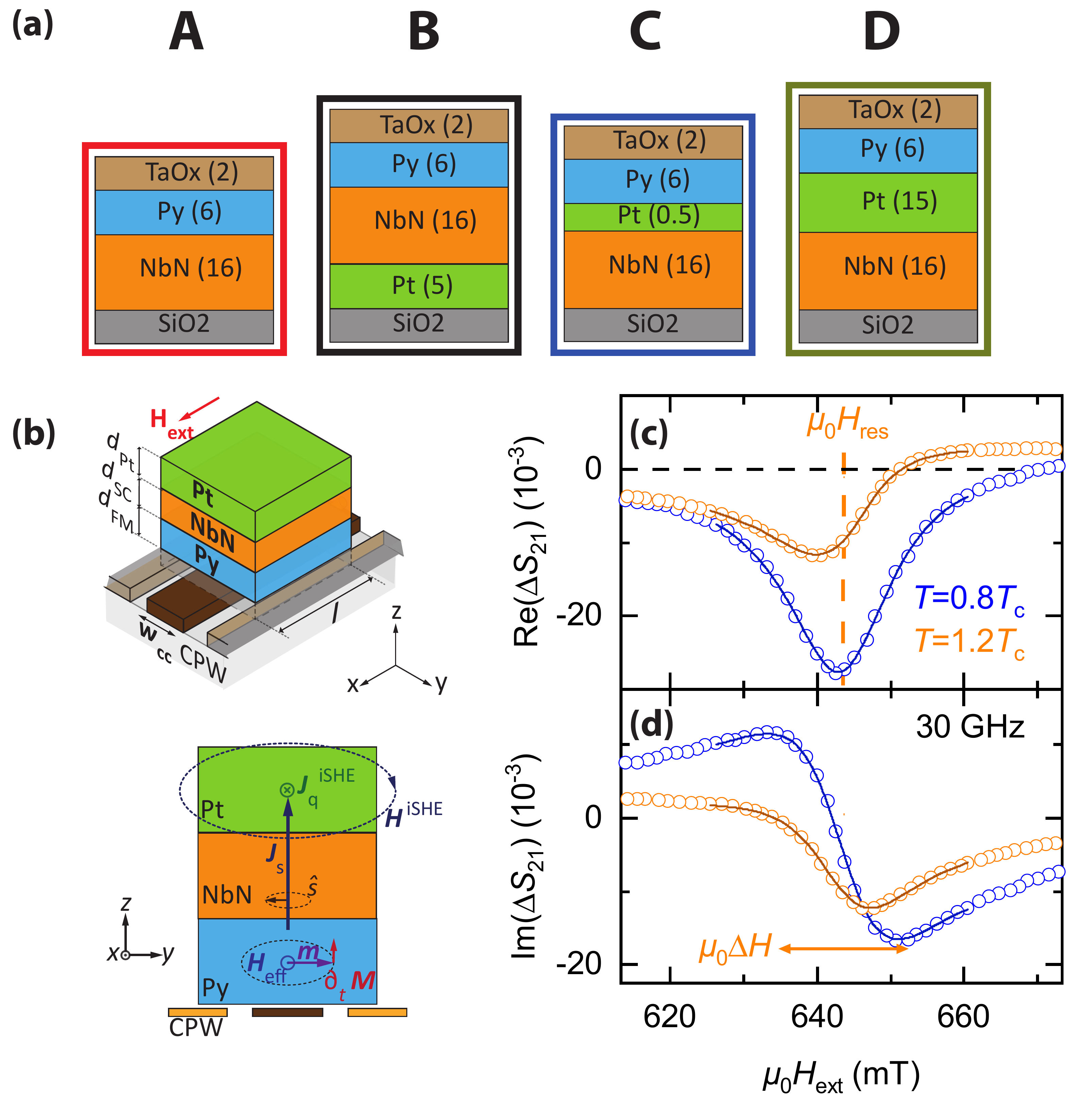}}
	\caption{(a): Layer stack of samples investigated in this publication. Numbers show the layer thicknesses in nm. For bbFMR, they are mounted face-down on top of a coplanar waveguide. (b): Sketch of the measurement geometry for in-plane bbFMR and schematic illustration for the generation of the charge current density $\boldsymbol{\mathrm{J}}_\mathrm{q}^{\mathrm{iSHE}}$ by ac iSHE. The ac flux $\boldsymbol{\mathrm{H}}^{\mathrm{iSHE}}$ generated by $\boldsymbol{\mathrm{J}}_\mathrm{q}^{\mathrm{iSHE}}$ is coupled into the CPW. In (c) and (d) the change in complex transmission is plotted versus the applied external field $\mu_0H_{\mathrm{ ext}}$ both slightly above and below $T_{\mathrm{c}}$ for sample B ($T_{\mathrm{c}}=9.0 $ K). Both FMR-amplitude and phase of the resonance exhibit clear changes in the SC state. The lines in (c) and (d) represent Polder susceptibility fits  to Eq. (S21). }
\label{Fig: series}
\end{figure}

Over the last decade, the field of superconducting spintronics  \cite{Sinha1982, Saxena2000, Sheikin2005, Jiang1995,Khusainov1997, Linder2015} has attracted increasing attention due to the novel and beneficial spin transport properties related to quasiparticles \cite{Inoue2017, Kato2019, Umeda2018, Morten2008} and spin-triplet Cooper pairs in superconductors (SCs) \cite{Linder2015, Keizer2006, Jeon2018}. Among those, charge-to-spin current interconversion and the associated spin-orbit torque effects allow for the control of magnetization and its dynamics. Investigations in this direction began with dc-transport experiments in SC lateral spin valve structures, which reported changes in spin signal and spin diffusion length $\lambda_{\mathrm{S}}$ \cite{Poli2008,Yang2010, Gu2002a} below  the superconducting transition temperature $T_{\mathrm{c}}$.\\Recent experiments \cite{Tserkovnyak2002, Bell2008, Yao2018, Jeon2018, Jeon2018a, Jeon2019,Jeon2019a, Jeon2019c} focused on the nonequilibrium magnetization dynamics of an FM layer adjacent to a SC film. Here, changes of the parameters describing magnetization dynamics below $T_{\mathrm{c}}$ provide insight into the spin injection in superconductors via spin pumping \cite{Bell2008, Yao2018, Jeon2018, Jeon2018a, Jeon2019, Jeon2019c}. Experiments analyzing the magnetization damping using ferromagnetic resonance (FMR) techniques in SC/FM hybrid systems close to $T_\mathrm{c}$ allowed to explore several phenomena, ranging from a monotonic reduction of spin pumping due to a freeze-out of quasiparticles (QP) \cite{Bell2008} over the manifestation of a coherence peak slightly below $T_{\mathrm{c}}$ \cite{Yao2018}, to the spin pumping mediated by spin-triplet Cooper pairs \cite{Jeon2018}. \\To clarify the origin of these competing results, we here present a systematic study of the magnetization dynamics of FM/SC-heterostructures as a function of temperature at and below $T_{\mathrm{c}}$.$\;$This includes the investigation of linear spin-orbit-torques present in these multilayers by employing broadband ferromagnetic resonance (bbFMR) experiments in combination with the phase sensitive detection of the microwave transmission signal. This approach allows us to simultaneously detect the electrical ac currents due to magnetization dynamics. These ac currents can arise due to inverse spin-orbit torques (iSOT) as well as classical electrodynamics (i.e. Faraday's law).$\;$For concise notation, we will quantify them in terms of $\sigma^\mathrm{SOT}$ in accordance with previous work \cite{Berger2018}. We are hence able to both quantify the impact of an adjacent SC film on the properties of the magnetization dynamics (e.g. FMR linewidth) and the field- and damping-like  $\sigma^\mathrm{SOT}$ in SC/FM heterostructures. For the latter, we expand the inductive coupling analysis reported in Ref. \cite{Berger2018} to account for superconducting layers. 
In this way, we establish a new powerful method to study nonequilibrium spin transport in SCs.\\For this study, we fabricate heterostructures based on NbN, Py, Pt, $\mathrm{TaO_x}$, which are  in-situ deposited on a thermally oxidized Si (100) substrate by dc magnetron sputtering (for more details on the deposition process see Supplemental Information \cite{Supplements}).\nocite{Horn1968, Hazra2016, Chand2009, Gubin2005, Gorter1934, Mhlschlegel1959, Chockalingam2008, Silva2017, Berger2018a, Maksymov2013, Maksymov2014, Phys2013, Rosa1908, Rojas-Sanchez2014, Gilmore2007, Flacke2019, Nembach2011, Gilbert2004, Polder1949, Ciovati2014,Imai2011,Onoda2006} While the thickness of NbN (16nm) and Py(6 nm) are constant, the Pt, acting as spin sink, is varied in thickness and position within the layer sequence (see Fig. \ref{Fig: series}(a)). $\mathrm{TaO_x}$ is used as the cap layer to prevent the oxidation of the heterostructure.
We perform bbFMR measurements in a cryogenic environment over a broad temperature range. A sketch of  the measurement geometry is shown in Fig. \ref{Fig: series}(b). The samples were mounted face-down onto a coplanar waveguide (CPW), of which we record the complex microwave transmission $S_{21}$ using a  vector-network analyzer (VNA). In particular, we set fixed microwave frequencies $f$ ($10 \mathrm{ GHz}\leq f\leq 36 \mathrm{ GHz}$) and measure as a function of the applied external magnetic field $H_{\mathrm{ext}}$ applied along the $\hat{\boldsymbol{\mathrm{e}}}_{\mathrm{x}}$-direction at a VNA output power of 1 mW. At such low power, all dynamics are in the linear regime. We first focus on sample B. For $T>T_{\mathrm{c}}$, the magnetization dynamics excited in the FM using FMR pump a spin current density $\boldsymbol{\mathrm{J}}_\mathrm{s}$ across the FM/SC interface ($T>T_{\mathrm{c}}$) and the SC into the adjacent Pt layer as illustrated in the inset of Fig \ref{Fig: series}(b). In the Pt layer, $\boldsymbol{\mathrm{J}}_\mathrm{s}$ is absorbed and converted into a charge current $\boldsymbol{\mathrm{J}}_\mathrm{q}$ via the inverse spin Hall effect (iSHE), where we assume a vanishing iSHE contributions from the SC-layer. This spin pumping effect manifests itself on the one hand as an additional contribution to the damping of the FMR as it represents an additional relaxation channel for angular momentum \cite{Tserkovnyak2002}. On the other hand, the ac magnetic field $H^{\mathrm{iSHE}}$ generated via the iSHE induced charge current $\boldsymbol{\mathrm{J}}_\mathrm{q}^{\mathrm{iSHE}}$ is inductively coupled into the CPW and thus can be detected by the VNA \cite{Berger2018}. Note that all experiments are performed for an ip-geometry ($\boldsymbol{\mathrm{H}}_{\mathrm{ ext}}\parallel\hat{\boldsymbol{\mathrm{e}}}_{\mathrm{x}}$) to avoid the  formation of vortices in the SC-state. Fig. \ref{Fig: series}(c) and (d) show data for the background-corrected change in transmission $\Delta S_{21}=\Delta S_{21}/\Delta S_{21}^0-1$ (see Eq. (S22)) in proximity to FMR for temperatures slightly above (orange) and below $T_{\mathrm{c}}$ (blue).
The raw data in Fig. \ref{Fig: series}(c) and (d) reveals that the SC transition significantly modifies the detected $\Delta S_{21}$-spectra. In the normal state ($T>T_{\mathrm{c}}$), the asymmetric lineshape of Re($\Delta S_{21}$) matches that of a (Py/Pt)-sample in Ref. \cite{Berger2018}, where the inductive coupling between currents generated in a normal metal (NM) by iSOT and a CPW were investigated at room temperature. 
Below $T_{\mathrm{c}}$, the dip-like shape of Re($\Delta S_{21}$), indicative of the absence of spin-to-charge current conversion, is mostly restored. We also observe an enhanced FMR amplitude suggesting a modified inductive coupling $\tilde{L}$ between the sample and CPW in the SC state. We employ a data analysis procedure based on \cite{Berger2018} and extract the normalized inductance $\tilde{L}=L/\chi$ by inserting the raw data, fitted to Eq. (S21), into Eq. (S27) (see Supplemental material). By applying this method, we also derive the resonance field $H_{\mathrm{res}}$ and linewidth $\Delta H$. Following Berger et al. \cite{Berger2018}, the frequency-dependent inductance $\tilde{L}$ is modeled as 

\begin{align}
\begin{aligned}
\tilde{L}&=\tilde{L}_0+\tilde{L}_{\mathrm{j}}\\
&=\tilde{L}_0+f\cdot(\mathrm{Re}(\Delta\tilde{L}_{\mathrm{j}})\epsilon_{\mathrm{r}}(f)+i\cdot\mathrm{Im}(\Delta\tilde{L}_{\mathrm{j}}) \epsilon_{\mathrm{i}}(f)).
\label{eq: L-tilde}
\end{aligned}
\end{align}

\begin{figure}
	\centerline{\includegraphics[scale=1.0]{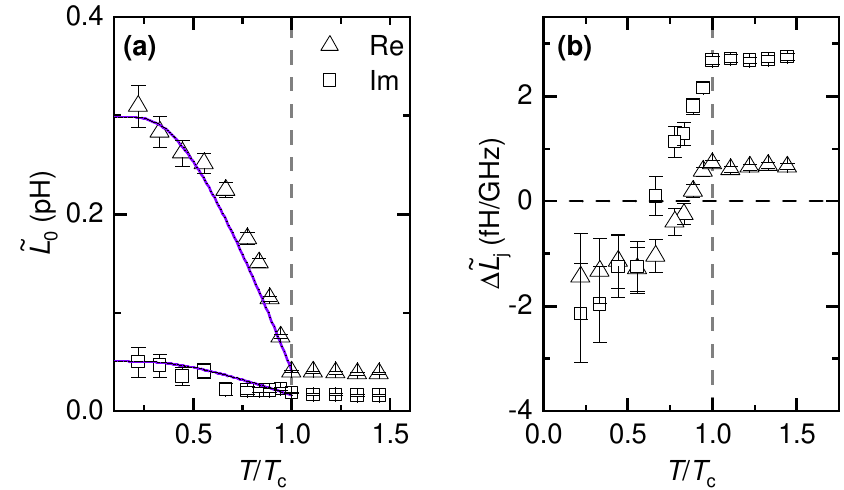}}
	\caption{Inductive coupling parameters for sample B with the stack sequence Pt (5 nm)/ NbN (16 nm)/ Py (6 nm)/ Ta (2 nm) plotted as a function of reduced temperature around $T_{\mathrm{c}}$. In (a), we plot the real (triangles) and imaginary (squares) parts of the inductive coupling offset $\tilde{L}_0$. The lines indicate the scaled superfluid condensate density $n_s$ following BCS-theory. In (b), we show real (triangles) and imaginary part (squares) of the complex linear frequency dependence $\Delta\tilde{L}_\mathrm{j}$ of the normalized inductance $\tilde{L}$. The error bars originate from fitting the extracted raw data for $\tilde{L}$ with Eq. (\ref{eq: L-tilde}). }
	\label{Fig: series2}
\end{figure}

 Above $T_{\mathrm{c}}$, $\tilde{L}_0$ is a measure of the coupling strength between the FM and the CPW and must be strictly real for $f\rightarrow0$ Hz, which is well met in our samples, while $\tilde{L}_\mathrm{j}$ accounts for the flux generated by ac charge currents in the adjacent NM/SC-layer. $\Delta\tilde{L}_{\mathrm{j}}$ denotes the linear frequency dependence of  $\tilde{L}_\mathrm{j}$ (see SI for a derivation of Eq. (\ref{eq: L-tilde})).  We further account for elliptical magnetization precession with the in-plane correction factors $\epsilon_{\mathrm{r}}(f)$ and $\epsilon_{\mathrm{i}}(f)$ for real and imaginary part, respectively (for details see Eq. (S41) in the supplemental material). We plot the extracted $\tilde{L}_0$ and  $\tilde{L}_{\mathrm{j}}$ as a function of reduced temperature for the Pt/NbN/Py-trilayer in Fig. \ref{Fig: series2}(a) and (b), respectively. Exemplary raw fitting data for $\tilde{L}(f)$ for all samples investigated are presented in the supplements.
Compared to $\tilde{L}_0$ at $T>T_\mathrm{c}$, we observe in Fig. \ref{Fig: series2}(a) a strongly enhanced coupling strength manifesting itself in a large $\mathrm{Re}(\tilde{L}_0)$ between the sample and CPW in the SC state that gradually increases with decreasing $T$. Simultaneously, $\mathrm{Im}(\tilde{L}_0)>0$  is observed for $T\mapsto 0$, while $\mathrm{Im}(\tilde{L}_0)\approx0$ for $T>T_{\mathrm{c}}$. Both observed effects are manifestations of superconductivity in NbN. In the SC state, image currents in the NbN layer mirror the external driving field of the CPW \cite{Schmidt} and keep the SC in the Mei{\ss}ner phase. Consequently, the sandwiched FM layer is driven from both sides with an enlarged net oscillatory driving field $\boldsymbol{\mathrm{h}}_{\mathrm{rf}}$, leading to the enhanced inductive coupling between sample and CPW. The nonzero $\mathrm{Im}(\tilde{L}_0)$ below $T_{\mathrm{c}}$ is attributed to the complex surface impedance $Z_{\mathrm{eff}}(\omega)$ in the SCs (see Supplemental material). \\We now turn to the  linear contribution to $\tilde{L}_\mathrm{j}$,  $\Delta\tilde{L}_\mathrm{j}$, shown in Fig. \ref{Fig: series2}(b). Note that $\tilde{L}_\mathrm{j}\neq0$ is obtained whenever electrical ac currents are generated in the samples by magnetization dynamics $\mathbf{m}(t)$. We find that both $\mathrm{Re}(\Delta\tilde{L}_\mathrm{j})$ (triangles, caused by currents in quadrature with $m_{\mathrm{y}}$) and $\mathrm{Im}(\Delta\tilde{L}_\mathrm{j})$ (squares, caused by currents in phase with  $m_{\mathrm{y}}$) change drastically just below $T_{\mathrm{c}}$. The real and imaginary part of $\tilde{L}_{\mathrm{j}}$ are attributed to the manifestation of field- and damping-like inverse current-induced torques (for an elaborate discussion see Supplemental Information). For a quantitative analysis of these effects, we extract the SOT conductivities $\sigma^{\mathrm{SOT}}$ from $\tilde{L}_{\mathrm{j}}$ by using the relation

\begin{align}
\tilde{L}_\mathrm{j}=C\cdot f\cdot[-\epsilon_{\mathrm{r}}(f)\sigma_{\mathrm{f}}+i\epsilon_{\mathrm{i}}(f)\sigma_{\mathrm{d}}].
	\label{eq: Lj}
\end{align}

\begin{figure}
	\centering
	\includegraphics[scale=1.0]{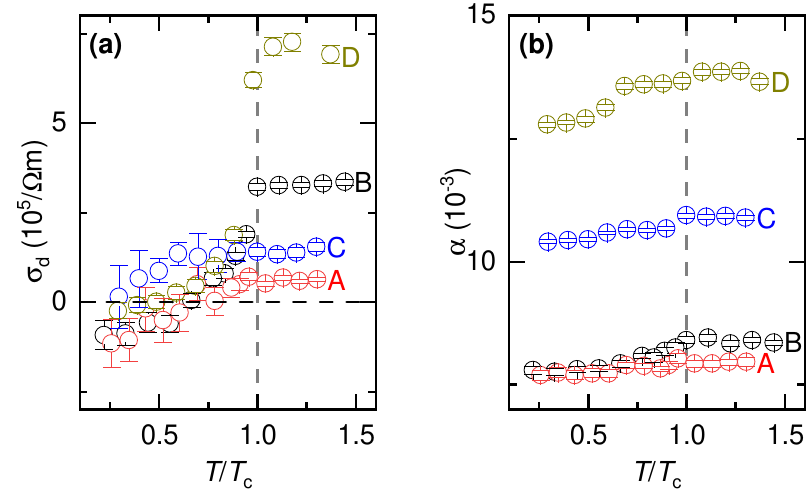}
	\caption{(a): Damping-like current-induced torque conductivity $\sigma_{\mathrm{d}}$ plotted as a function of reduced temperature. In the normal state, the samples containing Pt exhibit large positive  $\sigma_{\mathrm{d}}$ due to the iSHE in Pt. Below $T_{\mathrm{c}}$, all samples exhibit a very similar decay of $\sigma_{\mathrm{d}}$ with decreasing $T$ irrespective of Pt spin sink layer. (b): Temperature dependence of the Gilbert damping $\alpha$ in the SC state as a function of reduced temperature. The apparent decrease of $\alpha$ is due to the suppression of spin pumping into the superconductor due to the freeze-out of thermally excited quasiparticles. The error bars originate from fitting the extracted raw data for $\tilde{L}$ with Eq. (S43) (a) and Eq. (S17) for $\Delta H (f)$ (b).
	}
	\label{Fig: series3}
\end{figure}

\begin{figure}
	\centering
	\includegraphics[scale=1.0]{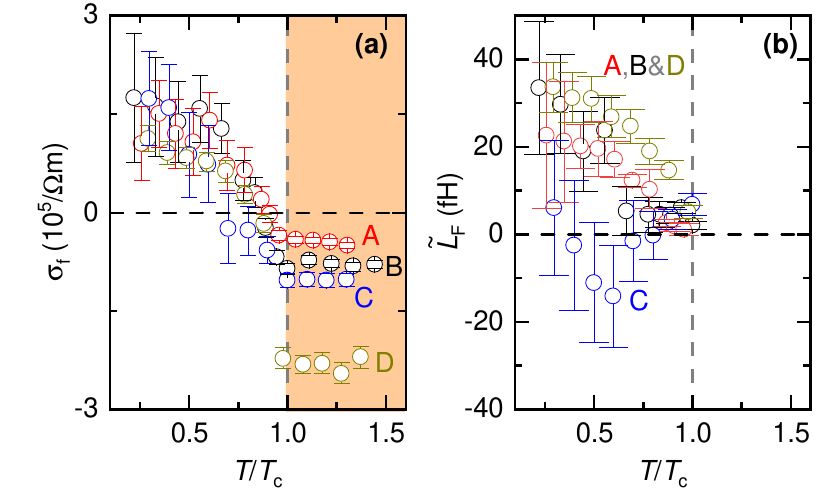}
	\caption{(a): Extracted field-like current-induced torque $\sigma_{\mathrm{f}}$ plotted as a function of reduced temperature. Due to a complex surface impedance $Z_{\mathrm{eff}}(\omega)$ in the SC, the Faraday contribution $\sigma_{\mathrm{f}}^{\mathrm{F}}$ creates an offset in  $\mathrm{Im}(\tilde{L})$ and only affects $\sigma_{\mathrm{f}}$ above  $T_{\mathrm{c}}$. We marked the corresponding temperature range with an orange background. In the SC state, we detect a substantial $\sigma_{\mathrm{f}}$. (b): Change in $\mathrm{Im}(\tilde{L})$ in the SC state due to SC Faraday currents.  The error bars in  (a) \& (b) originate from fitting the extracted raw data for $(\tilde{L})$ with Eq. (S43). 
	}
	\label{Fig: series4}
\end{figure}

Here, the indices f and d denote the field- and damping-like spin orbit torques, respectively. The field-like part $\sigma_{\mathrm{f}}$ also contains a Faraday contribution (magnetization dynamics in the FM induce dynamic charge currents in the adjacent NM), while $\sigma_{\mathrm{d}}$ is attributed to the iSHE. In our samples, the latter is dominated by the conversion of spin currents into charge currents in Pt. The proportionality constant $C$ in Eq. (\ref{eq: Lj}) is defined in the supplemental material (Eq. (S37)). Furthermore, below $T_{\mathrm{c}}$, we multiply an additional correction term to Eq. (\ref{eq: Lj}) to account for the altered net driving field strength $\boldsymbol{\mathrm{h}}_{\mathrm{rf}}$ (Eq. S43).
The inductive coupling strength $\tilde{L}_{\mathrm{j}}$ between NM and CPW is a linear function of $\sigma^{\mathrm{SOT}}$ \cite{Berger2018}. The damping-like and field-like current-induced torques $\sigma_{\mathrm{d}}$ and $\sigma_{\mathrm{f}}$ derived for the different samples are shown in Fig. \ref{Fig: series3} and \ref{Fig: series4}, respectively.
As shown in Fig. \ref{Fig: series3}(a), for $T>T_{\mathrm{c}}$, we observe large $\sigma_\mathrm{d}$ values for the samples B and D, as expected due to the iSHE in Pt. In contrast, sample A (NbN/Py bilayer) exhibits only small positive $\sigma_\mathrm{d}$ values, which we attribute to spin pumping into the unoxidized fraction of the $\mathrm{TaO_x}$-caplayer. For sample C, we determine a low $\sigma_{\mathrm{d}}$, indicating a reduction in the spin-to-charge current conversion due to the much lower thickness of the Pt layer. These results are in agreement with the expectations that Pt acts as a spin sink and efficiently converts spin- into charge currents.
 Below $T_\mathrm{c}$, $\sigma_{\mathrm{d}}$ rapidly decreases and eventually reaches a similar slightly negative value for all samples. There are two effects that affect $\sigma_{\mathrm{d}}$ in the SC state: First, there is a strongly modified spin transport carried by thermally excited QP and, second, there is a shunting effect by the SC condensate. For sample B, the spin current has to pass the SC layer to reach the Pt layer, such that altered spin transport properties in the SC play an important role in this case. The observed changes for the SC/Pt/FM-samples C and D, where the primary source of iSHE is the Pt-layer, can be well explained by the SC acting as a perfect electrical shunt, which reduces the charge current density $\mathrm{\boldsymbol{J}}_\mathrm{q}^{\mathrm{iSHE}}$ in Pt and as a result also $\sigma_{\mathrm{d}}$.
 A strong reduction in the detected magnitude of the iSHE in Pt due to shunting effects when brought into contact with highly conductive Cu has been reported in \cite{Yan2017, Kimura2007}. Here, we observe an analogue shunting effect in SC/Pt/FM-heterostructures, giving rise to the same strong suppression of the iSHE.
  The small negative values of $\sigma_{\mathrm{d}}$ in the SC state are consistent with the quasiparticle mediated inverse spin Hall effect (QMiSHE) in the superconductor \cite{Takahashi2012,Takahashi2008, Kontani2009, Takahashi2002}. The negative spin Hall angle \cite{Rogdakis2019,Wakamura2015} and the associated QMiSHE in NbN have recently been observed via non-local dc transport measurements \cite{Wakamura2015}. Remarkably, a diverging spin Hall angle resulting from intrinsic and side-jump contributions to the SHE compensates for a diminishing quasiparticle population with decreasing temperature (see supplemental information). A finite QMiSHE current is therefore expected in the superconducting state, as observed in Fig. \ref{Fig: series3}(a).
To support the validity of our approach and interpretation, we compare the damping-like $\sigma_{\mathrm{d}}$ to the extracted Gilbert damping $\alpha$ of Py, which is plotted as a function of reduced temperature in Fig. \ref{Fig: series3}(b). Due to spin pumping, $\alpha$ also directly probes  spin current transport in the heterostructure.
 The $\alpha$ value of sample A well matches literature values for the damping of Py thin films \cite{Bailey2001,Luo2014, Rantschler2005, Ghosh2011, Suraj2020} and hence serves as our reference sample in the absence of spin pumping. Consequently, larger $\alpha$ in the other samples originates from spin pumping into Pt. Thus, we observe substantial spin pumping contributions at all temperatures for samples C and D. In the SC regime, the $\alpha$ values of sample B approach the values of our NbN/Py-reference (sample A). This suggests a complete suppression of spin pumping into Pt at $T\ll T_{\mathrm{C}}$. In the NbN/Pt/Py-trilayers (samples C and D), the slight reduction of $\alpha$ below $T_{\mathrm{c}}$ is attributed to the blocking of spin currents at the NbN/Pt-interface. The $\alpha$ values of samples C and D remain substantially larger than those of sample A even for $T<T_\mathrm{c}$ because spin pumping into Pt is not affected by the superconducting NbN on the far interface.
The direct detection of dissipative spin currents via SOT in Fig. \ref{Fig: series3}(a) is thus consistent with their indirect detection via FMR damping $\alpha$ in \ref{Fig: series3}(b).\\Apart from $\sigma_{\mathrm{d}}$, originating from the iSHE, we can simultaneously extract $\sigma_{\mathrm{f}}$ generated by Faraday currents and field-like iSOT effects in the normal state. We plot the extracted $\sigma_{\mathrm{f}}$ for our samples in Fig. \ref{Fig: series4}(a).
 Above $T_{\mathrm{c}}$, we detect negative values for all samples attributed to Faraday currents in Pt, which scale with the Pt layer thickness in the respective sample. Below $T_\mathrm{c}$, all samples exhibit a small positive $\sigma_{\mathrm{f}}$ that gradually increases for decreasing $T$. Here, the observed behavior of all samples is nearly independent of the inclusion and position of the Pt layer. For $T<T_\mathrm{c}$, Faraday currents in the SC do not contribute to the slope of $\mathrm{Re}(\tilde{L})$ (and thus $\sigma_{\mathrm{f}}$) but generate an offset $\tilde{L}_\mathrm{F}$ in $\mathrm{Im}(\tilde{L})$ as illustrated in Fig. \ref{Fig: series4}(b). This explains the decrease of $|\sigma_{\mathrm{f}}|$ for $T<T_{\mathrm{c}}$ with simultaneous increase in $\tilde{L}_{\mathrm{F}}$ for all samples. Faraday currents are still present below $T_{\mathrm{c}}$, but now manifest in $\tilde{L}_{\mathrm{F}}$, because the superconductor now acts as an inductor and not a resistor (see Supplemental information Eq. (S45)).  The change in  $\mathrm{Im}(\tilde{L}_0)$ due to Faraday currents in the SC is similar in all samples except for sample C, where pronounced noise hinders the exact extraction of $\mathrm{Im}(\tilde{L}_0)$. Moreover, the positive $\sigma_{\mathrm{f}}$ for $T<T_{\mathrm{c}}$ in Fig. \ref{Fig: series4}(a) has the opposite sign compared to that expected from Faraday currents. $\sigma_{\mathrm{f}}$ for $T\mapsto 0$ also cannot be attributed to the inverse Rashba-Edelstein effect \cite{Bychkov1984a, Edelstein1990} at the SC/FM-interface, as the observed phenomenon is seemingly independent of the material adjacent to the SC and does not require direct contact between SC and FM. Having ruled out the common sources of field-like current-induced torques, we can merely speculate about its origin in our samples.
Potential candidates to generate $\sigma_{\mathrm{f}}$ in these samples include the coherent motion of vortices in an rf-field \cite{Dobrovolskiy2018, Awad2011} as well as the impact of Mei{\ss}ner screening currents on the magnetization dynamics due to triplet superconductivity \cite{Jeon2019} or non equilibrium effects \cite{Ouassou2020}.\\In summary, we adapt the method outlined in \cite{Berger2018} to detect the manifestation of substantial field-like  and the reduction of damping-like current-induced torques in FM/SC hybrids below $T_{\mathrm{c}}$.$\;$Our observations with respect to the damping-like current-induced torques are consistent with a shunting effect of the SC and quasiparticle-mediated iSHE in NbN. In particular, we establish a complementary detection technique for the QMiSHE, which  corroborates a recent report \cite{Wakamura2015}.$\;$The Gilbert damping of the FM layer $\alpha(T)$ demonstrates that spin-current transport through the SC in FM/SC/Pt-heterostructures is blocked below $T_{\mathrm{c}}$, while spin-pumping in FM/Pt/SC layers is only weakly affected by the superconducting transition. The sizable field-like current-induced torque below $T_{\mathrm{c}}$ does not originate from Faraday currents or interfacial iSOTs. This as yet unexplained observation raises interesting questions regarding the theoretical understanding of spin current transport in magnet/superconductor hybrids.\nocite{Horn1968, Hazra2016, Chand2009, Gubin2005, Gorter1934, Mhlschlegel1959, Chockalingam2008, Silva2017, Berger2018a, Maksymov2013, Maksymov2014, Phys2013, Rosa1908, Rojas-Sanchez2014, Gilmore2007, Flacke2019, Nembach2011, Gilbert2004, Polder1949, Ciovati2014,Imai2011,Onoda2006}
\section{Acknowledgments}
We acknowledge financial support by the Deutsche
Forschungsgemeinschaft (DFG, German Research Foundation) via WE5386/4-1, WE5386/5-1 and Germany’s Excellence Strategy EXC-2111-390814868 and the Research Council of Norway through its Centers of Excellence funding scheme, project 262633, ``QuSpin''.
\begin{acknowledgments}

\end{acknowledgments}



\bibliography{library}

\end{document}